\documentstyle[epsfig, aps,12pt,manuscript,amsbsy]{revtex}
\global\firstfigfalse
\tighten
\tightenlines

\begin{document}
\newcommand{\beq}{\begin{equation}}
\newcommand{\eeq}{\end{equation}}
\newcommand{\beqa}{\begin{eqnarray}}
\newcommand{\eeqa}{\end{eqnarray}}
\newcommand{\fr}{\frac}
\draft
\preprint{INJE-TP-03-03,hep-th/0303034}

\title{Dilatonic tachyon in  Kerr-de Sitter space}

\author{Y.S. Myung\footnote{Email-address :
ysmyung@physics.inje.ac.kr} and H. W. Lee\footnote{Email-address :
hwlee@physics.inje.ac.kr} }
\address{Relativity Research Center and School of Computer Aided Science\\
Inje University, Gimhae 621-749, Korea}

\maketitle

\begin{abstract}
We study the dynamical behavior of the dilaton in the background
of three-dimensional Kerr-de Sitter space which is  inspired from
the low-energy string effective action. The Kerr-de Sitter space
describes the gravitational background of a point particle whose
mass and spin are given by $1-M$ and $J$ and  its curvature radius
is given by $\ell$. In order to study the  propagation of metric,
dilaton, and Kalb-Ramond two-form, we perform the perturbation
analysis in the southern diamond of  Kerr-de Sitter space
including a conical singularity. It reveals a mixing between the
dilaton and other unphysical  fields.  Introducing a gauge-fixing,
we  can disentangle this mixing completely and obtain one
decoupled dilaton equation. However it turns out to be a tachyon
with $m^2=-8/\ell^2$. We compute the absorption cross section for
the dilatonic tachyon  to extract information on the cosmological
horizon of Kerr-de Sitter space. It is found that there is no
absorption of the dilatonic tachyon in the presence of the
cosmological horizon of Kerr-de Sitter space.
\end{abstract}

\newpage
\section {introduction}
Recently an accelerating universe has proposed to be a way
to interpret the astronomical data of supernova\cite{Per,CDS,Gar}.
The inflation is employed to solve the cosmological flatness and
horizon puzzles arisen in the standard cosmology.
Combining this observation with the need of inflation
 leads to that our universe approaches de Sitter
geometries in both the infinite past and the infinite future\cite{Wit,HKS,FKMP}. Hence it is
very important to study the nature of de Sitter (dS) space and the
dS/CFT correspondence\cite{BOU}.
 However,
there exist  difficulties in studying de Sitter space.
First there is no spatial infinity and   global timelike
Killing vector.  Thus it is not easy to define the conserved  quantities including  mass,
charge and angular momentum appeared in asymptotically  de Sitter space.
Second the dS solution is absent from string theories and thus we
do not   have a definite example  to test the dS/CFT correspondence.
Third it is hard to define  the $S$-matrix because of the
presence of the cosmological horizon.

Accordingly most of works on  de Sitter space were concentrated on
a free scalar propagation and its quantization\cite{STR,BMS,SV,YS,MLEE}. Also the
bulk-boundary relation for the free scalar was introduced to study the
dS/CFT correspondence\cite{AWLQ}. Hence it is important to find a model which
can accommodate  the de Sitter space solution. In this work we
introduce an interesting de Sitter model which is motivated from the
low-energy string action in (2+1)-dimensions\cite{Hor}. This model includes a
nontrivial scalar so-called the dilaton\footnote{Previously we are interested
in the Kerr-anti de Sitter space like the BTZ black hole\cite{BTZ}.
In the BTZ black hole and the three-dimensional black string,
the role of the dilaton was discussed in ref.\cite{flux}.}.
 Recently  we used the
dilaton to investigate the nature of the cosmological horizon in
de Sitter space\cite{LMY}.

It is well known that the cosmological horizon is very similar
to the event horizon in the sense that one can define its
thermodynamic quantities of a temperature and an entropy using the same way as is done for
the black hole. Two important quantities to understand the black hole
are the Bekenstein-Hawking entropy and the absorption cross
section (greybody factor). The former specifies   intrinsic
property of the black hole itself, while the latter relates to the
effect of spacetime curvature surrounding the black hole.
We emphasized here the greybody factor for
the black hole arises as a consequence of scattering off the
gravitational potential surrounding the event horizon\cite{grey1}. For example,
the low-energy $s$-wave greybody factor for a massless scalar has a
universality such that it
is equal to the area of the horizon for all spherically symmetric
 black holes\cite{grey2}. This can be obtained  by solving the wave
 equation explicitly.  The entropy for the cosmological horizon was
  discussed in literature\cite{entropy1,entropy2,shan}. However,
 there exist
 a few of the wave equation approaches to find the greybody factor  for the
 cosmological horizon\cite{YS,MLEE,LMY}.
 A similar work for the four-dimensional
 Schwarzschild-de Sitter black hole appeared  in\cite{KOY} but it
 focused mainly on obtaining the temperature of  the eternal  black hole.
Also the absorption rate for the four-dimensional Kerr-de Sitter
black hole was discussed in\cite{STU}.

In this paper we compute the absorption cross section of the
dilaton in the background of  three-dimensional Kerr-de Sitter
space with the cosmological horizon. For this purpose we first
confine  the wave equation only to the southern diamond where the
time evolution of  waves is properly defined even though it
contains a conical singularity at $r=0$. We mention that the
Kerr-de Sitter as a quotient of de Sitter space \cite{BMS} is not
appropriate for studying the dilaton wave propagation in the
southern diamond because the time coordinate is periodic in the
picture. This is useful for the study of thermal states and dS/CFT
correspondence. Hence  we study the perturbation in the background
of Kerr-de Sitter space itself. First we obtain approximate
solutions at $r=\pm i r_{(-)},0,r_+$, where $r=\pm i r_{(-)}$ are
complex locations and $r=r_+$ is the location of the cosmological
horizon. Then we solve the wave equation explicitly  to find the
greybody factor by transforming it into a hypergeometric equation.

The organization of this paper is as follows. In section II we
briefly review our model inspired from the low-energy string action and
its  Kerr-de Sitter solution. We
introduce
the perturbation to study   the propagation of fields in the
background of Kerr-de Sitter space in section III.
In section IV we solve the wave equation for the dilatoic tachyon.
In section V we calculate  the absorption cross section
 of the dilatonic tachyon explicitly.
Finally we discuss our results in section VI.

\section{Kerr-de Sitter solution}

\begin{figure}[t!]
\epsfig{file=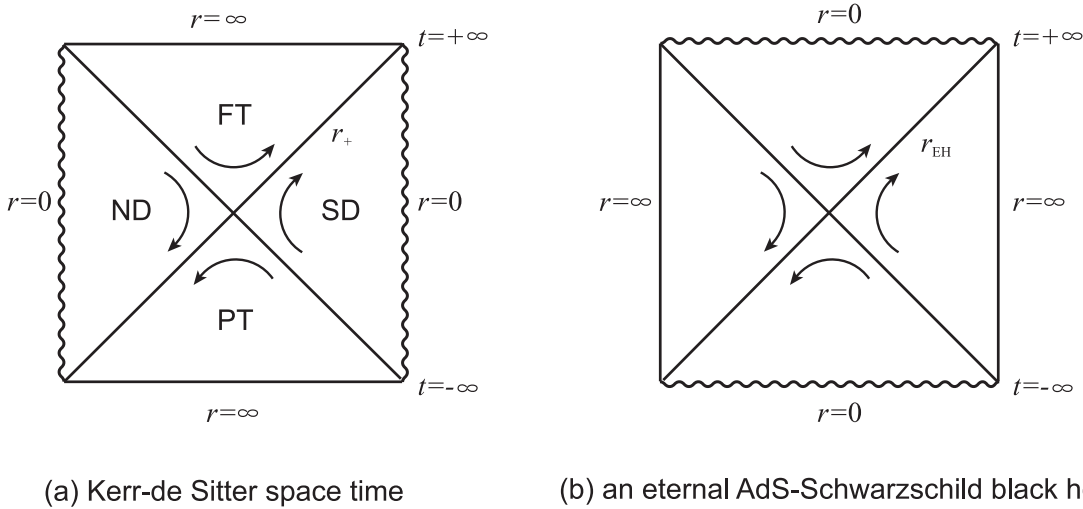,width=0.9\textwidth} \caption{The
upper/lower horizontal boundaries are future null infinity (${\cal
I}^+$)/past null infinity (${\cal I}^-$) of $r=\infty$. For
AdS-black hole, the horizontal lines belong to spacelike
singularity of $r=0$. The vertical lines are spacelike singularity
of $r=0$ for Kerr-de Sitter spacetime, while these are timelike
infinity of $r=\infty$ for AdS-black hole. For Kerr-de Sitter
spacetime, we call two regions with $0 \le r \le 1$ as northern
diamond (ND) and southern diamond (SD). An observer at $r=0$ is
surrounded by a cosmological horizon $r=r_+$. Two regions with $1
\le r \le \infty$ are called future triangle (FT) including ${\cal
I}^+$ and past triangle (PT) including ${\cal I}^-$. The correct
wave propagation is possible only in the southern diamond because
a timelike Killing vector $\partial /
\partial t$ is future-directed only in this region. The
arrow($\rightarrow$) is the flow direction generated by the
timelike Killing vector. Two of Kerr-de Sitter and AdS-black hole
do not have a global timelike Killing vector.} \label{fig1}
\end{figure}

We start with the low-energy string action in string frame\cite{Hor}
\begin{equation}
S_{l-s} = \int d^3 x \sqrt{-g} e^{\Phi}
   \big \{ R + (\nabla \Phi)^2 + {8 \over k} - {1 \over 12} H^2  \big \},
\label{action}
\end{equation}
where $\Phi$ is the dilaton, $H_{\mu\nu\rho}=3 \partial_{[\mu}B_{\nu\rho]}$
is the Kalb-Ramond field, and $k$ is related to the cosmological constant.
This action was widely used for studying the BTZ black hole (Kerr-anti de Sitter solution)
and
the black string\cite{flux}.
Although $k$ was originally proposed to be positive for the anti-de Sitter physics,
 we  choose it
to be negative  by analytic continuation of $\ell \to i\ell$. Then
the above action  is suitable for studying  the field propagation  in de Sitter
space. Because the dS solution may be  absent from string
theories, the above action is considered as a toy model to study the de Sitter physics.
The equations of motion lead to
\begin{eqnarray}
R_{\mu\nu} - \nabla_\mu \nabla_\nu \Phi
-{1\over 4} H_{\mu \rho \sigma} H_\nu^{\rho \sigma} &=& 0,
\label{eq_graviton} \\
\nabla^2 \Phi + (\nabla \Phi)^2 - {8 \over k} - {1 \over 6} H^2   &=& 0,
\label{eq_scalar1}  \\
 \nabla_\mu H^{\mu \nu \rho} + (\nabla_\mu \Phi) H^{\mu \nu \rho} &=& 0.
\label{eq_anti}
\end{eqnarray}
The Kerr-de Sitter  solution to
Eqs.(\ref{eq_graviton})-(\ref{eq_anti}) is found to be\cite{BMS,LMY}
\begin{eqnarray}
&& \bar H_{txr} =-i\fr{2r}{\ell}, ~~~~~~\bar
\Phi = 0,~~~~~~k=-2 \ell^2,
      \nonumber   \\
&& \bar g_{\mu\nu} =
 \left(  \begin{array}{ccc}  -(M - {r^2 / \ell^2}) & -{J / 2} & 0  \\
                             -{J / 2} & r^2 & 0  \\
    0 & 0 & f^{-2}
         \end{array}
 \right)
\label{bck_metric}
\end{eqnarray}
with the metric function $f^2 =M -r^2 / \ell^2  +J^2 / 4 r^2$.
The above Kerr-de Sitter solution can be obtained from the BTZ black hole
\cite{BTZ} by replacing both $M$ and $\ell^2$
by $-M$ and $-\ell^2$. One may worry about  the pure imaginary value of the Kalb-Ramond
 field in Eq.(\ref{bck_metric}). However, in the description of (Kerr)-de
 Sitter space in terms of a SL(2,C) Chern-Simons theory, the gauge
 field $A_\mu$ is usually given by the complex\cite{entropy1}. Hence
 we have
  no doubt of  choosing an imaginary value.
The metric $\bar g_{rr}$ is singular at $r=r_{\pm}$,
\begin{equation}
r_{\pm}^2 = {{M\ell^2} \over 2} \left \{ 1 \pm \left [
   1 + \left ( {J \over M\ell} \right )^2 \right ]^{1/2} \right \}
\label{horizon}
\end{equation}
with $M=(r_+^2 + r_-^2) / \ell^2=(r_+^2-r_{(-)}^2)/\ell^2$ and $ J=2 r_+r_{(-)} / \ell$.
Here $M$ is the mass of Kerr-de Sitter space with the cosmological
horizon at $r=r_+$ and $J$ is the angular momentum of it.
For convenience we introduce $r_{-}^2\equiv -r^2_{(-)}(r^2_{(-)}>0)$ due to $r^2_-<0$
in Kerr-de Sitter space. This means that $r_-=\pm i r_{(-)}$ becomes
purely imaginary and thus there exist only a cosmological horizon at $r=r_+$.
This point contrasts to that of the BTZ black hole because the
latter has two real horizons called the event(outer) horizon and
Cauchy(inner) horizon. As is shown in Fig.~\ref{fig1}, the Kerr-de Sitter solution
is more similar to the eternal AdS-Schwarzschild black hole than the BTZ
black hole.
There is no black hole horizon
for three-dimensional Kerr-de Sitter space because the black hole degenerates to a
conical singularity at the origin $r=0$. This space corresponds to
the gravitational background of a point particle at $r=0$ in de Sitter space whose mass and
spin are given by ${\cal M}$ and $J$, and  the  curvature radius of
de Sitter space is given by
$\ell$.  Although this singularity gives rise to
some difficulties to analyze the wave equation in the southern diamond (SD) of Kerr-de Sitter
space, we introduce an observer within SD for our purpose.
 In this work we mainly consider the
cosmological horizon  with interest.
For reference, we list the particle mass ${\cal M}$, Hawking temperature $T_c$,
area of cosmological horizon ${\cal A}_c$, and  angular velocity at the horizon
$\Omega_c$ as
\begin{equation}
{\cal M}=1-M,~~T_c = (r_+^2 + r_{(-)}^2) / 2 \pi \ell^2 r_+,
~~{\cal A}_c = 2 \pi r_+,
~~\Omega_c = J / 2 r_+^2
\label{temp}
\end{equation}
which are measured by an observer at $r=\infty$.
For $M=1,J=0$ case, one
finds de Sitter solution which gives us with $r_c=\ell$ and $r_-=0$
\begin{equation}
{\cal M}=0,~~T_c = 1 / 2 \pi \ell,
~~{\cal A}_c = 2 \pi \ell,
~~\Omega_c =0.
\label{tempd}
\end{equation}

\section{ Perturbation around Kerr-de Sitter solution}

To study the propagation of all fields in  Kerr-de Sitter space specifically,
 we introduce
the small perturbation fields\cite{flux}
\begin{equation}
H_{t \phi r} = \bar H_{t \phi r} + {\cal H}_{t \phi r},~~~
\Phi = 0 + \varphi,~~~
g_{\mu\nu} = \bar g_{\mu\nu} + h_{\mu\nu}
\label{per}
\end{equation}
around the background solution of
Eq.(\ref{bck_metric}). Here ${\cal H}_{t \phi r}=\partial_{[t}{\cal B}_{\phi
r]}$.
For convenience,
we introduce the notation
${\hat h}_{\mu\nu} = h_{\mu\nu}- {\bar g}_{\mu\nu} h/2$ with
$h= h^\rho_{~\rho}$.
And then one needs to linearize Eqs.(\ref{eq_graviton})-(\ref{eq_anti})
to obtain
\begin{eqnarray}
  \delta R_{\mu\nu} (h)
- \bar \nabla_\mu \bar \nabla_\nu \varphi
- {1 \over 2} \bar H_{\mu \rho \sigma} {\cal H}_\nu^{~ \rho \sigma}
+ {1 \over 2} \bar H_{\mu \rho \sigma} \bar H_{\nu\alpha}^{~~\sigma}
h^{\rho \alpha} &=& 0,
\label{lin_graviton} \\
 \bar \nabla^2 \varphi
- {1 \over 6} \Big \{ 2 \bar H_{\mu \rho \sigma}
                      {\cal H}^{\mu \rho \sigma}
     - 3 \bar H_{\mu \rho \sigma}   \bar H^{\alpha \rho \sigma} h^\mu_\alpha
                                    \Big \} &=& 0,
\label{lin_scalar} \\
   \bar \nabla_\mu  {\cal H}^{\mu \nu \rho}
- ( \bar \nabla_\mu h_\beta^{~\nu}) {\bar H}^{\mu\beta\rho}
+ (\bar \nabla_\mu h_\beta^{~\rho}) {\bar H}^{\mu\beta\nu}
- (\bar \nabla_\mu {\hat h}_{~\alpha}^{\mu}) {\bar H}^{\alpha\nu\rho}
      + (\partial_\mu \varphi) \bar H^{\mu \nu \rho}
 &=& 0,
\label{lin_anti}
\end{eqnarray}
where the Lichnerowicz operator $\delta R_{\mu\nu}(h)$
is given by
\begin{eqnarray}
&&\delta R_{\mu\nu} = -{1 \over 2} \bar \nabla^2 h_{\mu\nu}
 +{\bar R}_{\sigma ( \nu} h^\sigma_{~\mu )}
 -{\bar R}_{\sigma \mu\rho\nu} h^{\sigma\rho}
 + \bar \nabla_{( \nu} \bar \nabla_{|\rho|} {\hat h}^\rho_{~\mu)}.
\label{delR}
\end{eqnarray}
These belong to the bare perturbation equations.  It is desirable to examine whether
we  choose the physical perturbation by introducing  a gauge which can simplify
Eqs.(\ref{lin_graviton})-(\ref{lin_anti}) significantly.
For this purpose we wish to count the physical degrees of freedom first.
A symmetric
traceless tensor has components of D(D+1)/2--1 in D-dimensions.  D of them
are eliminated by the gauge condition.  Also D--1 are
eliminated from our freedom to take
further residual gauge transformations.
Thus gravitational degrees of freedom are D(D+1)/2--1--D--(D--1)=D(D--3)/2.
In three dimensions we have no
propagating degrees of freedom for metric fluctuation $h_{\mu\nu}$.
Also the Kalb-Romond two-form ${\cal B}_{\mu\nu}$ has no physical degrees of freedom for D=3.
Hence the physical degree of freedom in the Kerr-de Sitter solution
is just the dilaton $\varphi$.

Considering the $t$ and  $\phi$-symmetries of the background
spacetime Eq.(\ref{bck_metric}),
we can decompose $h_{\mu\nu}$ into frequency ($\omega$) and angular ($\mu=0,1,2,\cdots$)
modes in these
variables
\begin{equation}
h_{\mu\nu}(t,\phi,r) = e^{-i \omega t} e^{i\mu \phi}H_{\mu\nu}(r).
\label{ptr_metric}
\end{equation}
For simplicity, one chooses the same perturbation as in Eq.(\ref{ptr_metric})
 for Kalb-Ramond field and dilaton as
\begin{eqnarray}
{\cal H}_{t \phi r}(t,\phi,r) &&= \bar H_{t \phi r} {\cal H}(t,\phi,r)
 =\bar H_{t \phi r} e^{-i \omega t} e^{i \mu \phi} \tilde{\cal H}(r),
\label{ptr_anti} \\
\varphi(t,\phi,r)&&=e^{-i \omega t} e^{i \mu \phi} \tilde\varphi(r).
\label{ptr_scalar}
\end{eqnarray}
We stress again that this choice is possible only for D=3 because
$h_{\mu\nu}$ and ${\cal H}_{\mu\nu\rho}$
are redundant fields. Since the dilaton is only a propagating mode, we
focus on  the dilaton equation (\ref{lin_scalar}).
Eq.(\ref{lin_graviton}) is irrelevant to our analysis because
it gives us a redundant relation.
Eq.(\ref{lin_scalar}) can be rewritten as
\begin{eqnarray}
\bar \nabla^2 \varphi
+{4 \over l^2} (h - 2 {\cal H} )
=0.
\label{eq_scalar}
\end{eqnarray}
We wish  to decouple $h$ and ${\cal H}$ from the above $\varphi$-equation by
making use of gauge-fixing and the Kalb-Ramond equation.
If we start with full six degrees of freedom of Eq.(\ref{ptr_metric}), we
should choose a gauge.
Conventionally, we choose the
harmonic  gauge ($\bar \nabla_\mu {\hat h}^{\mu\rho} = 0$) to
describe the propagation of gravitons in D$>$3 dimensions\cite{wein}.
A mixing between the dilaton and other fields of $h,{\cal H}$  can be disentangled
 with the
harmonic gauge condition. Here we wish to introduce the
dilaton gauge
($\bar\nabla_\mu \hat h^{\mu \rho}=h^{\mu \nu} \Gamma^\rho_{\mu \nu} $) for
simplicity.
Actually this gauge was designed for the study of the dilaton propagation\cite{dilga}.
We attempt to disentangle the last term in Eq.(\ref{eq_scalar}) by
using both the dilaton gauge and
Kalb-Ramond equation (\ref{lin_anti}).
Each component ($\rho=t,\phi,r$) of the  dilaton gauge condition gives rise to
\begin{eqnarray}
t&:& (\partial_r + {1 \over r} ) h^{tr}
- i \omega h^{tt} + i \mu h^{t \phi}
+ {1 \over 2} i \omega h \bar g^{tt} - {1 \over 2}i \mu h \bar g^{t\phi} = 0,
\label{eq_gauge_t}  \\
\phi&:& (\partial_r + {1 \over r} ) h^{\phi r}
- i \omega h^{\phi t} + i \mu h^{\phi \phi}
+ {1 \over 2}i \omega h \bar g^{\phi t} - {1 \over 2}i \mu h \bar g^{\phi \phi} = 0,
\label{eq_gauge_x}  \\
r&:& (\partial_r + {1 \over r} ) h^{rr}
- i \omega h^{rt} + i \mu h^{r \phi}
- {1 \over 2} (\partial_r h) \bar g^{rr} = 0.
\label{eq_gauge_r}
\end{eqnarray}
And each component ($\nu,\rho$) of the Kalb-Ramond equation (\ref{lin_anti}) leads to
\begin{eqnarray}
t\phi:&& -\partial_r (\varphi+{\cal H} -{h^t_{~t}} -{h^\phi_{~\phi}})
+ {1 \over rf^2}\left (-M +{3 r^2 \over \ell^2} +
                     {J^2 \over 4r^2}\right ) h^r_{~r}
+i\omega h^t_{~r} -i \mu h^\phi_{~r}=0,
\label{eq_anti_tx} \\
tr:&& -i \mu (\varphi+{\cal H} -h^t_{~t} -h^r_{~r})
- {1 \over r} h^r_{~\phi}
+2 f^2 h^\phi_{~r}
-\partial_r h^r_{~\phi} +i\omega h^t_{~\phi} =0,
\label{eq_anti_tr} \\
\phi r:&& -i\omega (\varphi+{\cal H} -h^\phi_{~\phi} -h^r_{~r})
           + {1 \over r} h^r_{~t}
-{2 rf^2 \over \ell^2} h^t_{~r}
+\partial_rh^r_{~t} +i \mu h^\phi_{~t} =0.
\label{eq_anti_xr}
\end{eqnarray}
Solving six equations (\ref{eq_gauge_t})-(\ref{eq_anti_xr}) simultaneously, one finds
one constraint only
\begin{equation}
\partial_\mu (2 \phi +2 {\cal H} - h ) =0, ~~\mu=t,\phi,r
\label{eq_anti_simple}
\end{equation}
which leads to
$h - 2 {\cal H} =2 \varphi$.
This means that $h$ and $ {\cal H}$ becomes  redundant  if one
follows
the perturbations along Eqs.(\ref{ptr_metric})-(\ref{ptr_scalar}).
It confirms that our counting for degrees of freedom is correct.
We note here that the harmonic gauge with the Kalb-Ramond equation (\ref{lin_anti})
leads to the same constraint as in Eq.(\ref{eq_anti_simple}).
As a result, Eq.(\ref{eq_scalar}) becomes a decoupled dilaton equation
\begin{equation}
\bar \nabla^2 \varphi + \fr{8}{\ell^2} \varphi=0
\label{deq}
\end{equation}
which can be rewritten explicitly as

\begin{eqnarray}
\left [ f^2 \partial_r^2
+ \left\{ {1 \over r} (\partial_r rf^2) \right\} \partial_r
-{{J \mu \omega} \over {r^2 f^2}}
+{\omega^2 \over f^2}
+{(-M+{r^2 /\ell^2)} \over r^2 f^2} j^2
\right ] \tilde \varphi
+{8 \over l^2}\tilde \varphi
=0,
\label{eq_decoupled}
\end{eqnarray}
It is noted from Eq.(\ref{deq}) that if the last term is absent,
it corresponds to the wave equation of a freely  massive scalar in Kerr-de Sitter space.
Comparing this dilaton equation with the freely massive scalar
equation in Kerr-de Sitter space
\begin{equation}
\bar \nabla^2 \varphi_m -m^2 \varphi_m=0,
\label{meq}
\end{equation}
 the dilaton propagates on  Kerr-de Sitter space
with the negative mass square of  $m^2_{\varphi}=-8/\ell^2$.
In Kerr-anti de Sitter space of the BTZ black hole, this is a good test field
with $m^2_{\varphi}=8/\ell^2$. However the dilaton turns out to be a
tachyon in the Kerr-de Sitter space background.
Although in anti de Sitter space, the physical role of the tachyon is
not clear, the dilatonic tachyon  plays an
important role in the study of the de Sitter physics\cite{LMY}.
Hence we have to do a further study for the tachyonic dilaton in
the Kerr-de Sitter background.

\section {solution to wave equation}

\begin{figure}[t!]
\epsfig{file=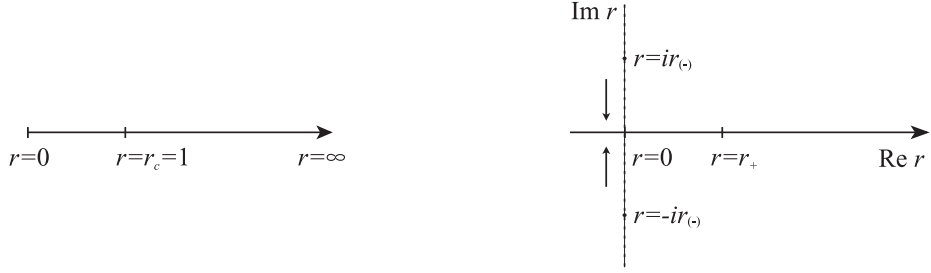,width=0.9\textwidth} \caption{In de Sitter
spacetime, $r=0$ is just a coordinate origin and $r=r_c=\ell$ is a
position of the cosmological horizon. On the other hand, for $J
\ne 0$ (Kerr-de Sitter space) $r=0$ becomes a conical singularity
where a spinning source exists. Instead a role of $r=0$ in de
Sitter space is split into $r=\pm i r_{(-)}$ ($r^2 = -r_{(-)}^2$)
in Kerr-de Sitter space. That is, if $J \rightarrow 0$, $\pm i
r_{(-)} \rightarrow 0$. Further $r=r_+$ is a position of the
cosmological horizon for $J \ne 0$. Hence, to solve the wave
equation explicitly, one introduces a coordinate $z=r^2/\ell^2
(0\le z \le 1)$ for de Sitter case, whereas one introduces $z =
{{r^2+r_{(-)}^2} \over {r_+^2 + r_{(-)}^2}}$( $0 \le z \le 1$) for
Kerr-de Sitter case. For $J \ne 0$, $r=0$ corresponds to $z_0
={{r_{(-)}^2} \over {r_+^2 + r_{(-)}^2}}$ located between $z=0$
and $z=1$. Even though this complex picture (b) seems to be
unphysical, it is useful for solving the wave equation in Kerr-de
Sitter space.} \label{fig2}
\end{figure}

It is not easy to solve the wave equation, Eq.(\ref{eq_decoupled}), of
the dilaton exactly
on the southern diamond  including  the cosmological horizon ($r=r_+$) and the origin ($r=0$).
 The main difficulty comes from the fact that the
black hole horizon degenerates to give a conical singularity at
$r=0$. In other words, the Kerr-de Sitter solution represents a spinning point mass
$1-M$ and spin $J$  within de Sitter space.
Before solving Eq.(\ref{eq_decoupled}), we present approximate
solutions around $r=0$, $r^2=-r_{(-)}^2$ and $r=r_+$.
For this purpose, see Fig.~\ref{fig2}.
Eq.(\ref{eq_decoupled}) can be
rewritten as
\begin{equation}
{\tilde \varphi}''
+ {1 \over f(r)} \left ( {f(r) \over r} + f'(r) \right ) {\tilde \varphi}'
+ {1 \over f(r)} \left [
-{{J \mu \omega} \over {r^2 f(r)}}
+{\omega^2 \over f(r)}
+{(-M+{r^2 /\ell^2)} \over r^2 f(r)} \mu^2 - m_\varphi^2
\right ] \tilde \varphi
=0.
\label{eq_norm}
\end{equation}
Each terms in Eq.(\ref{eq_norm}) also can be written in terms of
$r_+$ and $r_{(-)}$ as
\begin{equation}
{1 \over f(r)} \left ( {f(r) \over r} + f'(r) \right ) =
-{1 \over r} +{1\over {r_+ + r}} -{1\over {r_+ - r}}
+{1 \over {r+ i r_{(-)}}} +{1 \over {r- i r_{(-)}}} ,
\label{coef1_rprm}
\end{equation}
\begin{eqnarray}
&&{1 \over f(r)} \left [
-{{J \mu \omega} \over {r^2 f(r)}}
+{\omega^2 \over f(r)}
+{(-M+{r^2 /\ell^2)} \over r^2 f(r)} \mu^2 - m_\varphi^2
\right ] \nonumber \\
&&\hspace{10pt}=
{r^2\ell^4 \over {(r_+^2-r^2)^2(r^2+r_{(-)}^2)^2}}
\left [
\omega^2 r^2 -J \mu \omega +  \left (-M +{r^2 \over \ell^2}\right ) \mu^2
\right ] -
{r^2\ell^2m_\varphi^2 \over {(r_+^2-r^2)(r^2 + r_{(-)}^2)}} .
\label{coef0_rprm}
\end{eqnarray}

\subsection{Approximate Solution around $r=0$}
In this region the first term can be approximated as
\begin{equation}
\lim_{r \rightarrow 0} {1 \over f(r)} \left ( {f(r) \over r} + f'(r) \right ) \simeq
- {1 \over r} + {8M \over J^2} r + {\cal O}(r^3).
\label{coef1_r0}
\end{equation}
and the second one takes approximately
\begin{eqnarray}
&& \lim_{r \rightarrow 0} {1 \over f(r)} \left [
{\omega^2 \over f(r)}
-{{J \mu \omega} \over {r^2 f(r)}}
+{(-M+{r^2 /\ell^2)} \over r^2 f(r)} \mu^2 - m_\varphi^2
\right ]
\simeq
-C_0 r^2 + {\cal O}(r^4),
\label{coef0_r0}
\end{eqnarray}
where
\begin{equation}
C_0 = {16 \over J^4} \left [ J \omega \mu + M \mu^2
  + {{ m_\varphi^2 J^2} \over 4} \right ] .
\label{def-c0}
\end{equation}
Hence Eq.(\ref{eq_norm}) is reduced to
\begin{equation}
{\tilde \varphi_0}''
- {1 \over r} {\tilde \varphi_0}' - C_0 r^2 \tilde \varphi_0
=0,
\label{eq_approx_r0}
\end{equation}
which has the solution
\begin{equation}
{\tilde \varphi}_0 \simeq A_0 e^{-{{\sqrt{C_0}}\over 2 }r^2} +
  B_0 e^{{{\sqrt{C_0}}\over 2 }r^2}.
\label{sol_approx_r0}
\end{equation}
This reflects that $r=0$ is a conical singularity, where a
spinning source with $J \ne 0$ exists. In this sense  Kerr-de
Sitter space is different from de Sitter space. In the limit $J
\rightarrow 0$, this solution is meaningless because of
$\sqrt{C_0} \rightarrow \infty$.

\subsection{Approximate Solution around $r^2= -r_{(-)}^2$}
In this region, $r$  is pure imaginary. It can be understood that
$r$ is analytically extended here. We can split the limit $r^2
\rightarrow -r_{(-)}^2$ into $r \rightarrow i r_{(-)}$ and $r
\rightarrow - i r_{(-)}$. We consider the former because the
latter gives us the same result. The first term takes
\begin{equation}
\lim_{r \rightarrow i r_{(-)}} {1 \over f(r)} \left ( {f(r) \over r} + f'(r) \right ) \simeq
{1 \over {r- i r_{(-)}}}
\label{coef1_rmp}
\end{equation}
and the second one is given by
\begin{eqnarray}
&&\lim_{r \rightarrow i r_{(-)}} {1 \over f(r)} \left [
-{{J \mu \omega} \over {r^2 f(r)}}
+{\omega^2 \over f(r)}
+{(-M+{r^2 /\ell^2)} \over r^2 f(r)} \mu^2 - m_\varphi^2
\right ] \nonumber \\
&&\hspace{10pt} \simeq
- {1 \over 4}
 {1 \over {(r-i r_{(-)})^2}}{\ell^4 \over {(r_+^2+r_{(-)}^2)^2}}
\left[
\omega^2 r_{(-)}^2  + J \mu \omega + \left ( M + {r_{(-)}^2 \over \ell^2} \right ) \mu^2
\right]
\label{coef0_rmp}
\end{eqnarray}
Hence Eq.(\ref{eq_norm}) is approximated around $r=ir_{(-)}$ as
\begin{equation}
{\tilde \varphi_{-+}}''
+ {1 \over {r-ir_{(-)}}} {\tilde \varphi_{-+}}'
- {C_- \over {(r-ir_{(-)})^2}}{\tilde \varphi_{-+}}
\simeq 0,
\label{eq_approx_rmp}
\end{equation}
where
\begin{equation}
C_- = {\ell^4 \over {4(r_+^2+r_{(-)}^2)^2}}
\left[
J \mu \omega + \omega^2 r_{(-)}^2 + \left ( M + {r_{(-)}^2 \over \ell^2} \right ) \mu^2
\right] .
\label{cm}
\end{equation}
Here we have $C_- > 0$ and in the de Sitter limit, $C_- \to
 \mu^2/4$. Eq.(\ref{eq_approx_rmp}) has the solution
\begin{eqnarray}
{\tilde \varphi}_{-+} &=& A_{-+} (r-ir_{(-)})^{\sqrt{C_-}} +
B_{-+} (r-ir_{(-)})^{-\sqrt{C_-}} \nonumber \\
&=& A_{-+} e^{\sqrt{C_-} \ln (r-ir_{(-)})}
    + B_{-+} e^{-\sqrt{C_-} \ln (r-ir_{(-)})}.
\label{sol_approx_rmp}
\end{eqnarray}
Now we consider  the latter. In this case two terms are given
approximately by
\begin{equation}
\lim_{r \rightarrow -i r_{(-)}} {1 \over f(r)} \left ( {f(r) \over r} + f'(r) \right ) \simeq
{1 \over {r+ i r_{(-)}}}
\label{coef1_rmm}
\end{equation}
and
\begin{eqnarray}
&&\lim_{r \rightarrow - i r_{(-)}} {1 \over f(r)} \left [
-{{J \mu \omega} \over {r^2 f(r)}}
+{\omega^2 \over f(r)}
+{(-M+{r^2 /\ell^2)} \over r^2 f(r)} \mu^2 - m_\varphi^2
\right ]
\simeq
 - {C_- \over {(r+i r_{(-)})^2}}
\label{coef0_rmm}
\end{eqnarray}
Then Eq.(\ref{eq_norm}) is approximated around $r=-ir_{(-)}$ as
\begin{equation}
{\tilde \varphi_{--}}''
+ {1 \over {r+ir_{(-)}}} {\tilde \varphi_{--}}'
- {C_- \over {(r+ir_{(-)})^2}}{\tilde \varphi_{--}}
\simeq 0.
\label{eq_approx_rmm}
\end{equation}
The solution to Eq.(\ref{eq_approx_rmm}) is the same form as in
Eq.(\ref{sol_approx_rmp}),
\begin{eqnarray}
{\tilde \varphi}_{--} &=& A_{--} (r+ir_{(-)})^{\sqrt{C_-}} +
B_{--} (r+ir_{(-)})^{-\sqrt{C_-}} \nonumber \\
&=& A_{--} e^{\sqrt{C_-} \ln (r+ir_{(-)})}
    + B_{--} e^{-\sqrt{C_-} \ln (r+ir_{(-)})}.
\label{sol_approx_rmm}
\end{eqnarray}
Taking  the limit of $r_{(-)} \rightarrow 0$ ($J \rightarrow 0$),
 Eqs.(\ref{sol_approx_rmp}) and (\ref{sol_approx_rmm}) give the
same solution for consistency, which  implies that
\begin{eqnarray}
&& A_{-+} = A_{--} \equiv A_-, \nonumber \\
&& B_{-+} = B_{--} \equiv B_- .
\end{eqnarray}
Using these, we express Eqs.(\ref{sol_approx_rmp}) and (\ref{sol_approx_rmm}) as
\begin{eqnarray}
{\tilde \varphi}_{-+} &=& A_{-} (r-ir_{(-)})^{\sqrt{C_-}} +
B_{-} (r-ir_{(-)})^{-\sqrt{C_-}} \nonumber \\
&=& A_{-} e^{\sqrt{C_-} \ln (r-ir_{(-)})}
    + B_{-} e^{-\sqrt{C_-} \ln (r-ir_{(-)})},
\label{sol_approx_rmp1} \\
{\tilde \varphi}_{--} &=& A_{-} (r+ir_{(-)})^{\sqrt{C_-}} +
B_{-} (r+ir_{(-)})^{-\sqrt{C_-}} \nonumber \\
&=& A_{-} e^{\sqrt{C_-} \ln (r+ir_{(-)})}
    + B_{-} e^{-\sqrt{C_-} \ln (r+ir_{(-)})}.
\label{sol_approx_rmm1}
\end{eqnarray}
Note that if one takes the limit of $r_{(-)} \rightarrow 0$,
Eqs.(\ref{eq_approx_rmp}) and
(\ref{eq_approx_rmm}) becomes
\begin{equation}
{\tilde \varphi_{-0}}''
+ {1 \over r} {\tilde \varphi_{-0}}'
- {\mu^2 \over {4 M} }{1 \over r^2}{\tilde \varphi_{-0}}
=0,
\label{eq_approx_rm0}
\end{equation}
which has the Kerr-de Sitter solution at $r=0$ when $J=0$.
\begin{eqnarray}
{\tilde \varphi}_{-0} &=& A_{-} r^{\mu \over {2 \sqrt{M}}} +
B_{-} r^{-{\mu \over {2 \sqrt{M}}}} \nonumber \\
&=& A_{-} e^{{\mu \over {2 \sqrt{M}}} \ln r}
    + B_{-} e^{-{\mu \over {2 \sqrt{M}}} \ln r}.
\label{sol_approx_rm0}
\end{eqnarray}
This approximate solution is obtained by taking the limit $r_{(-)} \rightarrow 0$,
followed after $r\rightarrow 0$.
On the other hand, if we take the limit $r\rightarrow 0$, followed after
$r_{(-)} \rightarrow 0$, then Eq.(\ref{eq_norm}) becomes
\begin{equation}
\tilde \varphi_{0-}'' + {1 \over r} \tilde \varphi_{0-}'
- {\mu^2 \over M} {1 \over r^2} \tilde\varphi_{0-} = 0.
\label{eq_r0}
\end{equation}
This equation is actually different from Eq.(\ref{eq_approx_rm0}) and has
the de Sitter solution at $r=0$
\begin{eqnarray}
\tilde \varphi_{0-} &=&A_-'r^{{\mu} \over {\sqrt{M}}} +
                     B_-'r^{-{\mu} \over {\sqrt{M}}} \nonumber \\
&=& A_-'e^{{{\mu} \over {\sqrt{M}}}\ln r} +
                     B_-'e^{{-{\mu} \over {\sqrt{M}}}\ln r}.
\label{sol_r0}
\end{eqnarray}
Hence the order of taking a limit is important to derive
the correct solution.

\subsection{Approximate Solution around $r= r_+$}
In this region two terms can obtained by using (\ref{coef1_rprm}) and
(\ref{coef0_rprm}) as
\begin{equation}
\lim_{r \rightarrow r_+} {1 \over f(r)} \left ( {f(r) \over r} + f'(r) \right ) \simeq
-{1 \over {r_+-r}}
\label{coef1_rp}
\end{equation}
and
\begin{eqnarray}
&&\lim_{r \rightarrow r_+} {1 \over f(r)} \left [
-{{J \mu \omega} \over {r^2 f(r)}}
+{\omega^2 \over f(r)}
+{(-M+{r^2 /\ell^2)} \over r^2 f(r)} \mu^2 - m_\varphi^2
\right ] \nonumber \\
&&\hspace{10pt}\simeq
{1 \over 4} {1 \over {(r_+-r)^2}}{\ell^4 \over {(r_+^2+r_{(-)}^2)^2}}
\left[
 \omega^2 r_+^2 -J \mu \omega  + \left ( -M + {r_+^2 \over \ell^2} \right ) \mu^2
\right]
\label{coef0_rp}
\end{eqnarray}
Hence Eq.(\ref{eq_norm}) is approximated around $r=r_+$ as
\begin{equation}
{\tilde \varphi_+}''
- {1 \over {r_+-r}} {\tilde \varphi_+}'
+ {C_+ \over {(r_+-r)^2}}{\tilde \varphi_+}
\simeq 0,
\label{eq_approx_rp}
\end{equation}
where
\begin{equation}
C_+ = {\ell^4 \over {4(r_+^2+r_{(-)}^2)^2}}
\left[
\omega^2 r_+^2 -J \mu \omega  + \left ( -M + {r_+^2 \over \ell^2} \right ) \mu^2
\right] .
\label{cp}
\end{equation}
We assume that $C_+ > 0$ for our purpose. This is valid because in
the de Sitter limit of $J\rightarrow 0$, $r_+\rightarrow \ell$, $M
\rightarrow 1$, one finds that $C_+ \rightarrow \omega^2
\ell^2/4>0$. Then Eq.(\ref{eq_approx_rp}) has the travelling wave
solution
\begin{eqnarray}
{\tilde \varphi}_+ &=& A_+ (r_+-r)^{i\sqrt{C_+}} + B_+ (r_+-r)^{-i\sqrt{C_+}} \nonumber \\
&=& A_+ e^{i\sqrt{C_+} \ln (r_+-r)}
    + B_+ e^{-i\sqrt{C_+} \ln (r_+-r)}.
\label{sol_approx_rp}
\end{eqnarray}

\subsection{Exact Solution for the whole region}
Now we can convert Eq.(\ref{eq_decoupled}) to a hypergeometric equation by introducing
variable $z \equiv (r^2+ r_{(-)}^2)/(r_+^2 + r_{(-)}^2)$ as (see Fig.~\ref{fig2})
\begin{equation}
{{d^2 \tilde \varphi} \over {dz^2}} +
\left (
{1 \over z} - {1 \over {1 - z}}
\right ) {{d \tilde \varphi} \over {dz}} -
\left (
{C_- \over z} - {C_+ \over {1-z}} + {{m_\varphi^2 \ell^2} \over 4}
\right ) {{\tilde \varphi} \over {z(1-z)}} = 0 ,
\label{eq-hyper1}
\end{equation}
where $m_\varphi^2 \ell^2 = -8$, $C_+$ and $C_-$ are defined in
Eqs.(\ref{cp}) and (\ref{cm}).
Note that $z \to 0$ corresponds to $r \to \pm i r_{(-)}$.
With a new function $\tilde \varphi = z^\alpha (1-z)^\beta \bar \varphi$,
Eq.(\ref{eq-hyper1}) can be transformed into
\begin{eqnarray}
z(1-z) \bar\varphi'' &+&
\left [ 1 + 2 \alpha - ( 2 + 2 \alpha + 2 \beta ) \right ] \bar\varphi' \nonumber \\
&-& \left [
(\alpha + \beta)(\alpha + \beta + 1) + {{m^2 \ell^2} \over 4}
- {{\alpha^2 - C_-} \over z} - {{\beta^2+C_+} \over {1-z}}
\right ] \bar\varphi = 0 .
\label{eq-hyper2}
\end{eqnarray}
This equation leads to the standard hypergeometric equation if we
choose $\alpha = \sqrt{C_-}$ and $\beta = i \sqrt{C_+}$. Other
choices of $\alpha$ and $\beta$ are irrelevant to our
purpose\cite{MLEE}. Then a normalizable solution near $z=0$ is
\begin{equation}
\tilde\varphi = E_0 z^{\sqrt{C_-}} (1-z)^{i\sqrt{C_+}} F(a,b,c;z),
\label{sol-exact}
\end{equation}
with
\begin{eqnarray}
a &=& {1 \over 2} \left (
1 + 2 \sqrt{C_-} + 2 i \sqrt{C_+} + \sqrt{1-m^2 \ell^2}
\right )
\label{a-def} \\
b &=& {1 \over 2} \left (
1 + 2 \sqrt{C_-} + 2 i \sqrt{C_+} - \sqrt{1-m^2 \ell^2}
\right )
\label{b-def} \\
c &=& 1 + 2 \sqrt{C_-}
\label{c-def}
\end{eqnarray}
and an arbitrary constant $E_0$. If we take the de Sitter limit
($J \to 0$, $M \rightarrow 1$, $r_+\rightarrow \ell$), the above
quantities are reduced to\cite{MLEE}
\begin{eqnarray}
&&\sqrt{C_-} \to {\mu \over {2 }}, ~~ \sqrt{C_+} \to {{\omega \ell} \over {2}}
\label{ab-limit} \\
&& a \to {1 \over 2} (1 + \mu + i \omega \ell + \sqrt{1-m_\varphi^2 \ell^2}),
\nonumber \\
&& b \to {1 \over 2} (1+ \mu + i \omega \ell - \sqrt{1-m_\varphi^2 \ell^2}),
\label{abc-limit} \\
&& c \to 1 +  \mu .
\nonumber
\end{eqnarray}

We transform Eq.(\ref{sol-exact}) to a region around $z=1$ using the relation
between the hypergeometric functions
\begin{eqnarray}
&&F(a,b,c;z)=
\fr{\Gamma(c)\Gamma(c-a-b)}{\Gamma(c-a)\Gamma(c-b)}
F(a,b,a+b-c+1;1-z)\\ \nonumber
&& ~~+ \fr{\Gamma(c)\Gamma(a+b-c)}{\Gamma(a)\Gamma(b)}(1-z)^{c-a-b} F(c-a,c-b,-a-b+c+1;1-z).
\label{f_relation}
\end{eqnarray}
Using $1-z \simeq 2r_+(r_+-r)/(r_+^2 + r_{(-)}^2) $$= 2r_+/(r_+^2 + r_{(-)}^2)e^{\ln(r_+-r)}$,
one finds from Eqs.(\ref{sol-exact}) and (\ref{sol_approx_rp}) the following form :
\begin{equation}
\tilde \varphi_{z\rightarrow 1} \equiv \tilde\varphi_{in} + \tilde\varphi_{out}
= A_+e^{i\sqrt{C_+} \ln(r_+-r)} + B_+e^{-i\sqrt{C_+} \ln(r_+-r)},
\label{sol_z1}
\end{equation}
where $A_+$ and $B_+$ are determined as
\begin{eqnarray}
A_+ &=& E_0 \left ( {{2 r_+} \over {r_+^2 + r_{(-)}^2}} \right )^{i \sqrt{C_+}}
{{\Gamma(c) \Gamma(c-a-b)} \over {\Gamma(c-a)\Gamma(c-b)}}
\label{def_ap} \\
B_+ &=& E_0 \left ( {{2 r_+} \over {r_+^2 + r_{(-)}^2}} \right )^{-i \sqrt{C_+}}
{{\Gamma(c) \Gamma(a+b-c)} \over {\Gamma(a)\Gamma(b)}}
\label{def_bp}
\end{eqnarray}
Note that $A_+ = B_+^*$ if $E_0$ is chosen to be  real. Because we
find $|A_+/B_+|=1$, two travelling waves
$(\tilde\varphi_{in}e^{-i\omega t},~\tilde
\varphi_{out}e^{-i\omega t})$ have amplitudes of the same
magnitude. The absolute square of the amplitude of the tachyonic
dilaton must be proportional to the flux. As a result we conclude
that an outgoing $(\leftarrow)$ wave is reflected back to give an
ingoing $(\rightarrow)$ wave by the potential.  This
interpretation is  in accordance with the classical picture of
what the particle goes on\cite{Wich}. Actually there is no wave
propagating truly toward a conical singulaity.

\section{No absorption cross section}

The absorption coefficient by the cosmological horizon is defined by
 the ratio of the outgoing flux at $r=0$ to the outgoing flux at $r=r_+$
 as
\beq {\cal A} = \fr{ {\cal F}_{out}(r=r_+)}{{\cal F}_{out}(r=0)}.
\label{5eq1} \eeq Since the wave function near $r=0$ is real (see
Eq.(\ref{sol_approx_r0})), its flux should be  zero (${\cal
F}_{out}(r=0)=0$). Also ${\cal F}_{out}(r=r_+)$ can be calculated
as \beq {\cal F}_{out}(r=r_+) = {{2 \pi} \over i} \left . \left [
\tilde\varphi_{out}^* (r_+-r) \partial_r \tilde\varphi_{out} -
\tilde\varphi_{out} (r_+-r) \partial_r \tilde\varphi_{out}^*
\right ] \right |_{r=r_+} = 4 \pi \sqrt{C_+} \left | B_+ \right
|^2. \label{flux-def} \eeq On the other hand, the ingoing flux at
$r=r_+$ takes the form \beq {\cal F}_{in}(r=r_+) = {{2 \pi} \over
i} \left . \left [ \tilde\varphi_{in}^* (r_+-r) \partial_r
\tilde\varphi_{in} - \tilde\varphi_{in} (r_+-r) \partial_r
\tilde\varphi_{in}^* \right ] \right |_{r=r_+} = 4 \pi \sqrt{C_+}
\left | A_+ \right |^2. \label{flux-in} \eeq Here we find ${\cal
F}_{out}(r=r_+) = {\cal F}_{in}(r=r_+)$ because of $|B_+| =
|A_+|$. This means that there is no absorption of the  tachyonic
dialton in Kerr-de Sitter space. This is obvious because  there is
no wave propagating truly toward a conical singulaity. This can be
also checked by  the zero flux of ${\cal F}_{out}(r=0)=0$.  As a
result, the absorption cross section $\sigma_{\rm abs}$ defined
${\cal A}/\omega$ in three dimensions is zero,
\begin{equation}
\sigma_{\rm abs} = 0
 \label{abs-x-section}
\end{equation}
even if one gets ${\cal F}_{out}(r=0)=0$. One finds the same
situation if one uses $r=\pm i r_{(-)}$ instead of $r=0$ to
calculate the denominator of Eq.{(\ref{5eq1}).

\section{discussion}
We study the wave equation of the tachyonic dilaton
 which propagates on the southern diamond of
three-dimensional Kerr-de Sitter space. We wish to point out that
this space is somewhat different from de Sitter space because it
contains a conical singularity at $r=0$. First we find approximate
solutions at $r=\pm i r_{(-)},~0,~r_+$ to the wave equation.  We
compute the absorption cross section
 to investigate  its cosmological horizon  by transforming the
 wave equation into the standard hypergeometric equation.
By analogy of the quantum mechanics of the wave scattering under the potential step,
 it turns out that there is no
absorption of the tachyonic dilaton in Kerr-de Sitter space in the
semiclassical approach. This means that Kerr-de Sitter space is
usually stable and in thermal equilibrium, unlike the black hole.
The cosmological horizon not only emits radiation but also absorbs
that previously emitted by itself at the same rate, keeping the
curvature radius $\ell$ of Kerr-de Sitter space fixed. This can be
proved by the relation of ${\cal F}_{out}(r=r_+) ={\cal
F}_{in}(r=r_+)$ and ${\cal F}_{out/in}(r=0)=0$. This exactly
coincides with the wave propagation of the energy $E$ under the
potential step with $0<E<V_0$ which shows the classical picture of
what the particle goes on\cite{Wich}. Here we find a nature of the
eternal Kerr-de Sitter horizon\cite{DKS}, which means that its
cosmological constant $\Lambda=1/l^2$ remains unchanged, as like
the eternal AdS-black hole\cite{Mal}.

Finally we remark on a few of results. 1) A conical singularity at
$r=0$ is not so important to derive our conclusion of no
absorption in Kerr-de Sitter space. In Kerr-de Sitter space,  a
role of the coordinate origin $r=0$ in de Sitter space is replaced
by not $r=0$ but $r=\pm r_{(-)}$. 2) In the black hole study we
usually avoid a naked singularity by making use of the cosmic
censorship hypothesis. That is, the working region is from the
event horizon to the infinity. However, in the de Sitter physics,
we cannot avoid a singularity because our interesting region is
the southern diamond (SD) in Fig.1 which includes a conical
singulaity. 3) In this work we deal directly with a conical
singularity to find the absorption cross section of the dilatonic
tachyon. 4) Either the presence of a conical singularity (Kerr-de
Sitter space) or the absence of a conical singularity (de Sitter
space) gives us the same zero absorption cross section. This
implies that a conical singularity does not affect on the
absorption cross section of any scalar wave with the mass square
$m^2$ including the dilatonic tachyon with
$m^2_{\varphi}=-8/\ell^2$. In this sense the mass of a scalar wave
does not play an important role in propagation of Kerr-de Sitter
space.

\section*{Acknowledgements}
Y.S. was supported in part by  KOSEF, Project No.
R02-2002-000-00028-0. H.W. was in part supported by KOSEF,
Astrophysical Research Center for the Structure and Evolution of
the Cosmos.


\begin{references}
\bibitem{Per} S. Perlmutter et al.(Supernova Cosmology Project), Astrophys. J.
{\bf 483}, 565(1997)[astro-ph/9608192].

\bibitem{CDS}R. R. Caldwell, R. Dave, and  P. J. Steinhard, Phys. Rev. Lett.
{\bf 80}, 1582(1998)[astro-ph/9708069].

\bibitem{Gar}P. M. Garnavich et al.,  Astrophys. J. {\bf 509}, 74(1998)[astro-ph/9806396].

\bibitem{Wit} E.Witten, ``Quantum Gravity in de Sitter Space",
hep-th/0106109.

\bibitem{HKS}S. Hellerman, N. Kaloper, and L. Susskind, JHEP {\bf 0106}, 003(2001)[hep-th/0104180].

\bibitem{FKMP}  W. Fischler, A. Kashani-Poor,
R. McNees, and  S. Paban, JHEP {\bf 0107}, 003 (2001)[hep-th/0104181].

\bibitem{BOU} R.Bousso, JHEP {\bf 0011}, 038 (2000)[hep-th/0010252];
R. Bousso, JHEP {\bf 0104}, 035 (2001)[hep-th/0012052];
S. Nojiri and D. Odintsov, Phys.Lett. {\bf B519}, 145 (2001)[
hep-th/0106191]; D. Klemm, ``Some Aspects of the de Sitter/CFT
correspondence", hep-th/0106247 ;
M. Spradlin, A. Strominger, and A. Volovich, ``Les
Houches Lectures on De Sitter Space", hep-th/0110007;
S. Cacciatori and D. Klemm, ``The Asymptotic
Dynamics of de Sitter Gravity in three Dimensions", hep-th/0110031;
 A. C. Petkou and G. Siopsis, ``dS/CFT correspondence
on a brane", hep-th/0111085.
V. Balasubramanian, J. de Boer, and D. Minic,
``Mass, Entropy and Holography in Asymptotically de Sitter
Spaces", hep-th/0110108;  R. G. Cai, Y. S. Myung, and Y. Z. Zhang, ``Check of
the Mass Bound Conjecture in de Sitter Space", hep-th/0110234;
 Y. S. Myung, Mod. Phys. Lett.{\bf A16}, 2353(2001)[hep-th/0110123];
 R. G. Cai, ``Cardy-Verlinde Formula and
Asymptotically de Sitter Spaces", hep-th/0111093;
 A. M. Ghezelbach and R. B. Mann, JHEP {\bf 0201}, 005 (2002)[hep-th/0111217];
 M. Cvetic, S. Nojiri, and S.D. Odintsov,
``Black Hole Thermodynamics and Negative Entropy in deSitter and
Anti-deSitter Einstein-Gauss-Bonnet gravity", hep-th/0112045;
Y. S. Myung, ``Dynamic dS/CFT correspondence using the brane
cosmology", hep-th/0112140; R. G. Cai, ``Cardy-Verlinde Formula and
Thermodynamics of Black Holes in de Sitter Spaces",
hep-th/0112253; S. R. Das, ``Thermality in de Sitter and Holography",
hep-th/0202008;  F. Lebond, D. Marolf and R. C. Myers, ``Tall tales
from de Sitter space I: Renorlamization group flows",
hep-th/0202094; A. J. Medved, ``How Not to Construct an
Asymptotically de Sitter Universe", hep-th/0203191;
G. Siopsis, ``An ADS/DS Duality for a Scalar Particle",
hep-th/0203208;
T. R. Govindarajan, R. K. Kaul, and V. Suneeta, ``Quantum Gravity
on $dS_3$, hep-th/0203219;
C. Teitelboim, ``Gravitational Thermodynamics of Schwarzschild de
Sitter space", hep-th/ 0203258;
D. Klemm and L. Vanzo, ``De Sitter gravity and Liouville Theory",
hep-th/0203368.

\bibitem{STR} A. Strominger, JHEP {\bf 0110}, 034 (2001)[hep-th/0106113];
 A. J. Tolley and N. Turok, ``Quantization of the
massless minimally coupled scalar field and the dS/CFT
correspondence", hep-th/0108119.

\bibitem{BMS} R. Bousso, A. Maloney, and A. Strominger,
``Conformal vacua and entropy in de Sitter space", hep-th/0112218.


\bibitem{SV} M. Spradlin and A. Volovich, ``Vacuum states and the
S-matrix in dS/CFT", hep-th/0112223.

\bibitem{YS} Y. S. Myung, ``Absorption cross section in de Sitter
space", hep-th/0201176.

\bibitem{MLEE} Y. S. Myung and H. W. Lee, ``No absorption in de
Sitter space", hep-th/0302148.

\bibitem{AWLQ} Z. Chang and C.-B. Guan, ``Dynamics of Massive
Scalar Fields in dS Space and the dS/CFT Correspondence",
hep-th/0204014; E. Abdalla, B. Wang, A. Lima-Santos and W. G. Qiu,
``Support of dS/CFT correspondence from perturbations of three
dimensional spacetime", hep-th/0204030.


\bibitem{Hor}
G. Horowitz and D. Welch, Phys. Rev. Lett. {\bf 71}, 328(1993);
N. Kaloper, Phys. Rev. {\bf D48}, 2598(1993);
A. Ali and A. Kumar, Mod. Phys. Lett. {\bf A8}, 2045(1993).

\bibitem{BTZ}
M. Banados, C. Teitelboim and A. Zanelli, Phys. Rev. Lett.
{\bf 69}, 1849(1992).

\bibitem{flux} D. B. Birmingham, I. Sachs and S. Sen, Phys. Lett.
{\bf B413}, 281(1997); H.W. Lee, N. J. Kim, and Y. S. Myung,
Phys. Rev. {\bf D58}, 084022 (1988)[hep-th/9803080];
H.W. Lee, N. J. Kim, and Y. S. Myung,
Phys. Lett. {\bf B441}, 83(1988)[hep-th/9803227];
H.W. Lee and Y. S. Myung,
Phys. Rev. {\bf D58}, 104013 (1988)[hep-th/9804095].

\bibitem{LMY} H.W. Lee and Y. S. Myung,
Phys. Lett. {\bf B537}, 117(2002) [hep-th/0204083].

\bibitem{grey1} C. Callan, S. Gubser, I. Klebanov, and A.
Tseytlin, Nucl. Phys. {\bf B489}, 65(1997)[hep-th/9610172]; M. Karsnitz and I.
Klebanov, Phys. Rev. {\bf D56}, 2173(1997)[hep-th/9703216]; B. Kol and A.
Rajaraman, Phys. Rev. {\bf D56}, 983(1997)[hep-th/9608126];
M. Cvetic and F. Larsen, Nucl. Phys. {\bf B506},
107(1997)[hep-th/9706071].


\bibitem{grey2} A. Dhar, G. Mandal, and S. Wadia
,Phys. Lett. {\bf B388}, 51(1996) [hep-th/9605234];
 S. Das, G. Gibbons and S. Mathur, Phys. Rev. Lett. {\bf 78}, 417(1977)[hep-th/9609052].



\bibitem{entropy1}  M. I. Park, Phys.Lett. {\bf B440}, 275(1998)
 [hep-th/9806119]; M. Banados, T. Brotz and M. Ortiz,  Phys. Rev. {\bf D59},
 046002(1999)[hep-th/9807216].

\bibitem{entropy2}  W. T. Kim,
 Phys. Rev. {\bf D59}, 047503(1999)[hep-th/9810169]; F. Lin and Y. Wu,
 Phys. Lett. {\bf B453}, 222 (1999)[hep-th/9901147].

\bibitem{shan} S. Shankaranarayanan, ``Temperature and entropy of
Schwarzschild-de Sitter space-time", gr-qc/0301090.



\bibitem{KOY} W. T. Kim, J.J. Oh, and K. H.Yee,
``Scattering amplitudes and thermal temperatures of the
Schwarzschild-de Sitter black holes", hep-th/ 0201117.

\bibitem{STU} H. Suzuki, E. Takasugi, and H. Umetsu, Prog. Theor.
Phys. {\bf 103}, 103 (2000)[gr-qc/9911079].


\bibitem{wein} S. Weinberg, {\it Gravitation and Cosmology}
(Wiley, New York,1972), p.254.
\bibitem{dilga} H.W. Lee, N. J. Kim, Y. S. Myung, and J. Y. Kim,
Phys. Rev. {\bf D57}, 7361(1998)[hep-th/9801152].

\bibitem{Wich} E. H. Wichmann, {\it quantum physics-berkely
physics course 4} (Macgraw-hill, 1971, New York), p.280.
\bibitem{DKS} L. Dyson, M. Kleban, and L. Susskind, ``Disturbing
Implications of a Cosmological constant", hep-th/0208013.
\bibitem{Mal} J. Maldacena. ``Eternal black hole in Anti de
Sitter", hep-th/0106112.

\end{references}
\end{document}